\documentstyle[prb,aps,eqsecnum,preprint]{revtex}
\begin{document}
\draft
\title{COMPLETE WETTING IN THE THREE-DIMENSIONAL
TRANSVERSE ISING MODEL}
\author{A. B. Harris,$^1$ C. Micheletti,$^2$ and J. M. Yeomans$^2$}
\address{(1) Department of Physics,
University of Pennsylvania, Philadelphia, PA 19104-6396}
\address{(2) Theoretical Physics, Oxford University,
1 Keble Rd. Oxfored OX1 3NP, UK}
\date{\today}
\maketitle
\begin{abstract}
We consider a three-dimensional Ising model in a transverse magnetic
field, $h$ and a bulk field $H$. An interface is introduced by an
appropriate choice of boundary conditions. At the point $(H=0,h=0)$
spin configurations corresponding to  different positions of the
interface are degenerate. By studying the phase diagram near this
multiphase point using quantum-mechanical perturbation theory we show
that that quantum fluctuations, controlled by $h$, split the
multiphase degeneracy giving rise to an infinite sequence of layering
transitions.
\end{abstract}


\input{psfig}

\section{INTRODUCTION}

There is a considerable body of literature discussing the way in which
interfaces depin from surfaces[\onlinecite{unbinding}]. Of particular
interest to us here is the situation below the roughening transition
when the interface is smooth and can depin from the surface through a
series of first-order layering transitions.

This possibility was first pointed out by De Oliveira and
Griffiths[\onlinecite{DG78}] for a model in which the layering is driven
by the competition between a long-range bulk interaction and a surface
field. Here the layering transitions exist even at zero
temperature. Later Duxbury and Yeomans[\onlinecite{DY85}] showed that,
if the position of the interface relative to the surface was
degenerate at zero temperature, the degeneracy could be split by
thermal fluctuations giving an infinite series of layering transitions
at finite temperatures. The stable interface position is determined as
a balance between the binding effect of a bulk field and the entropic
advantage for the interface lying further from the surface.

The transverse Ising model [\onlinecite{RS73}] was first introduced by
de Gennes [\onlinecite{dG63}] in connection with ferro-electric materials.
Recently Henkel et.\ al.[\onlinecite{HH95}] have discussed the
behaviour of a domain wall in this system. The
interesting questions concern the effect of {\em quantum} fluctuations,
mediated by the transverse field, on the behaviour of the
interface. Their work considers one dimension where the interface is
rough. Very different behaviour is likely for a smooth
interface. Therefore in this paper we consider the behaviour of the
three-dimensional
transverse Ising model below the roughening temperature. We find that
a zero-temperature multiphase point can be split by quantum
fluctuations and that, for a non-zero transverse field, there are an
infinite number of stable positions for the interface as a bulk field
passes through zero.

The next section of the paper defines the model and gives a
qualitative discussion of its properties and our
approach. Quantitative details of the calculation, which is based on
quantum mechanical perturbation theory, are given in Sec. III.  Our
conclusions are summarized in Sec. IV.

\section{QUALITATIVE REMARKS}

\begin{figure}

\centerline{\psfig{figure=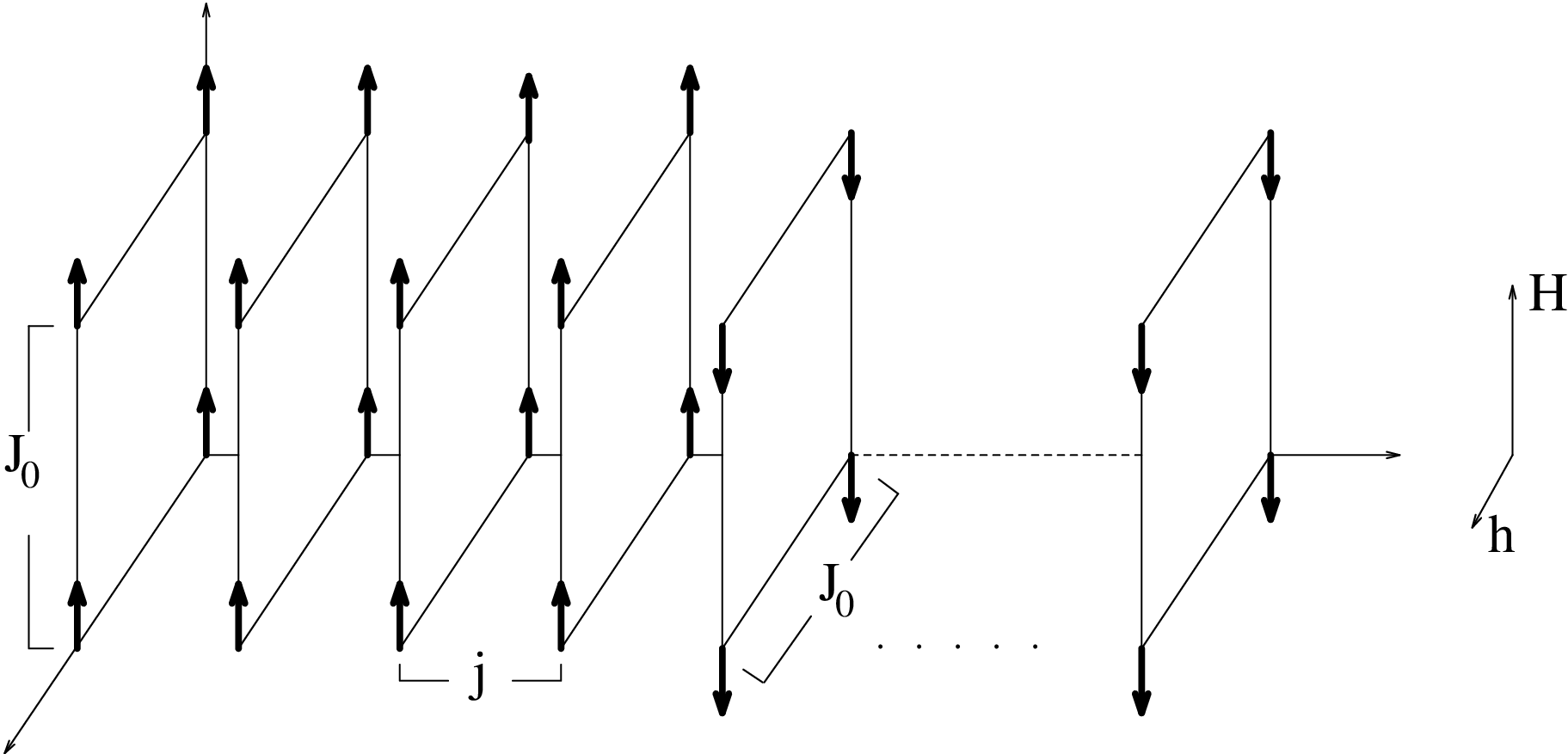,width=14.5cm}}
\vspace{0.5cm}
\caption{Schematic representation of the model introduced
in Eq. (\protect{\ref{MODEL}}). The leftmost plane is labeled $i=0$ and
the rightmost $i=L+1$.  The surface field $K$ is not shown.}
\label{f1}
\end{figure}

\begin{figure}

\centerline{\psfig{figure=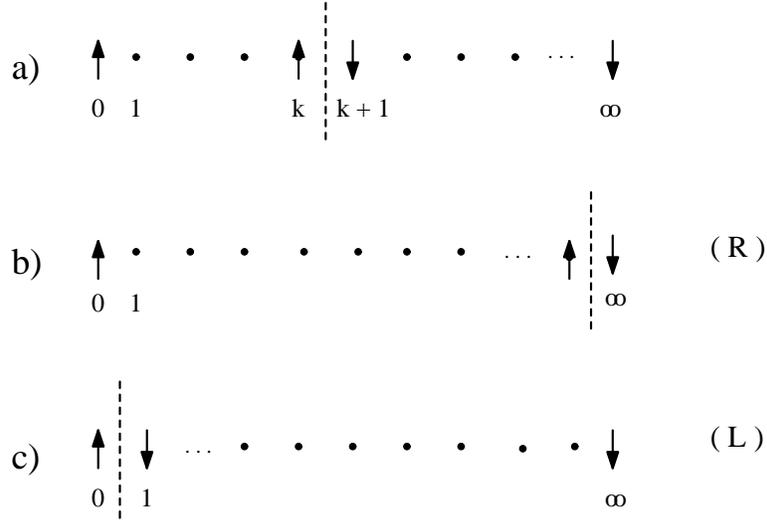,width=4.0in}}

\vspace{0.5cm}
\caption{a) Configuration of the spins when the interface is at
position $k$. Layer 0 is constrained to be up and the layer at infinity is
constrained to be down. In panels (b) and (c) we have represented
the configurations R and L.}
\label{f2}
\end{figure}

The Hamiltonian of the three-dimensional transverse Ising model we
shall consider is
\begin{eqnarray}
\label{MODEL}
{\cal H} = & - & J_0 \sum_{i=1,L} \sum_{\langle j,j'\rangle}
S_z (i,j) S_z(i,j')  - j \sum_{i=0,L} \sum_j
S_z(i,j) S_z(i+1,j) \nonumber \\
&  - &  \sum_{i=1,L} \sum_j
\biggl[ h  S_x (i,j) + H S_z(i,j) \biggr]
- K  \sum_j \biggl(  S_z (0,j) - S_z(L+1,j) \biggr) \ ,
\end{eqnarray}
where $i$ labels two-dimensional planes and $\langle j,
j^{\prime}\rangle$ nearest neighbours within a plane. The
parameter $K \rightarrow \infty$ is used to impose appropriate
boundary conditions, namely to fix the spins at the surface ($i=0$) to
be up and those in the last layer ($i=L+1$) to be down. These boundary
conditions will create a domain wall, or interface, in the system
separating layers of up and down spins (see Fig. \ref{f1}).
Our aim is to construct the phase diagram which gives the
position, $k$, of the interface (defined as in Fig. \ref{f2}a.)
as a function of the uniform field $H$ and the transverse field $h$.
We shall consider the limit
$L \rightarrow \infty$ and zero temperature where the interface is
flat.

As a first step in understanding the phase diagram we consider
the situation for $h=0$.  In this case it is clear that for
positive $H$, $k=\infty$ whereas for negative $H$, $k=0$.
We shall call these phases, which are illustrated in
Fig. \ref{f2}, R and L respectively.
For $H$=0 the energy is independent of $k$, so that
all interface positions are degenerate. It is
known that such a degeneracy can be lifted by either thermal
fluctuations[\onlinecite{Selke}] or by quantum
fluctuations[\onlinecite{HM95}].  In more general contexts this
removal of degeneracy has been referred to as ground state
selection[\onlinecite{CLH}] following the work of
Villain and Gordon[\onlinecite{JV}] and Shender[\onlinecite{EFS}].
Here we will consider the effect of quantum fluctuations.

There are two possibilities: in the first, quantum fluctuations
due to nonzero $h$ cause the transition from L to R to
be discontinuous with no intermediate states; in the
second this transition occurs through a sequence of
intermediate states in which $k$ increases monotonically.
This sequence can be
finite, so that there is a first-order transition from
a state with $k=k_{\rm max}$ to R, or it can be infinite.
In general, one expects the first possibility when the effective
interaction between the surface and the interface is attractive and
the second when this effective interaction is repulsive.
Our results indicate that quantum effects give rise to the second
possibility and that the sequence of layering transitions is probably
an infinite one.  It is easy to see that even when $h$ is nonzero,
the R phase is stable whenever $H$ is positive.  As we will show,
the stability of the L phase requires that $H < - C h^2 j/J_0^2$,
where $C$ is a constant.

To determine the interface position
we must calculate the ground state energy as a
function of $H$.  To do this we assume perfectly flat interfaces
and apply perturbation theory to calculate the energy of
the state when the interface is at position $k$.  If we were
dealing with a finite system, then perturbation theory
would introduce coupling in finite order between states
when the interface is at different positions.  However, this
tunneling effect disappears in the thermodynamic limit.
Our calculations are valid for
$h \ll J$.  We also impose the condition $j \ll J_0$,
although as long as we remain in the regime where the
interfaces are flat, this restriction is probably inessential.

The energy per layer spin $e(k)$ of
the state when the interface is at position $k$ can be written
\begin{equation}
\label{WALL}
e(k) = e_0(H) - kH + E_k(H) \ ,
\end{equation}
where $e_0(H)$ is the energy per layer spin for the $k=0$ phase.
The most important dependence on $H$ is in the term $-kH$.
$H$ also contributes to the energy
denominators that appear in perturbation theory, which are typically
of the form $2J_0+j+H$.
However perturbative contributions to $E_k(H)$, when expanded in
powers of $H$, lead to corrections which are of relative order
$H /J_0$ or smaller.  But since we will only be interested in
$H$ in the range $-Ch^2/J_0 < H < 0$, these corrections are
smaller than of relative order $(h^2 /J_0^2) \ll 1$, which we may ignore.
Therefore, in Eq. (\ref{WALL}), we may evaluate $E_k(H)$ at $H=0$.
We shall find that $E_k$ is a positive and decreasing function of $k$.
This result leads to an infinite sequence of phase transitions.  The
critical field separating the phase $k$ from $k+1$ is given by
\begin{equation}
\label{HEQ}
-H_k^* = E_k - E_{k+1} \equiv \Delta E_k \ ,
\end{equation}
where, as argued above, the leading order calculation
of $E_k$ can be obtained for $H=0$.

\section{CALCULATION OF $\Delta E_k$}

{}From Eq. (\ref{HEQ}) it is clear that we only need to keep track of
terms in the ground-state energy which depend on $k$.
In other words we need to ascertain how the corrections
to the ground--state energy which are perturbative in $h$
depend on the location of the interface.
For convenience we now transform to occupation number operators.
For spins that are up (down) we write $S_z(i,j) = 1/2-n_{i,j}$
($S_z(i,j)=-1/2+n_{i,j}$).  Also $S_x(i,j) = (a^\dagger_{i,j} + a_{i,j})/2$,
where the operator $a^\dagger_{i,j}$ ($a_{i,j}$)
creates (destroys) a Bose excitation at site $i$ in the $j$th layer
and $n_{i,j}=a_{i,j}a_{i,j}^\dagger$.
Strictly speaking we should not allow more than one excitation to exist
on a single site.  To enforce this restriction we include a term
of the form $\Lambda \sum_{i,j}  a_{i,j}^\dagger a_{i,j}^\dagger
a_{i,j} a_{i,j}$, where
$\Lambda \rightarrow \infty$.  Normally, it is difficult to
take full account of such a hard-core interaction.  As we will
see, we accommodate this constraint by never involving matrix
elements connecting to a state in which there is more than one
excitation at any site. Therefore, setting $H=0$, we are lead to
the following bosonic Hamiltonian, when the interface is at position $k$:
\begin{eqnarray}
{\cal H}_B = && E_0 +
\sum_{i=1}^\infty \sum_j \biggl( 2J_0 n_{i,j}
- (h/2)[a^\dagger_{i,j}+a_{i,j}] \biggr)
- \sum_{i=1}^\infty \sum_{\langle j,j'\rangle} J_0 n_{i,j} n_{i,j'}
+ \Lambda \sum_{i=1}^\infty \sum_j a_{i,j}^\dagger a_{i,j}^\dagger
a_{i,j} a_{i,j}
\nonumber \\ &&
-j\sum_{i=0}^\infty \sum_j [ - (1/2)(n_{i,j}+n_{i+1,j})
+ n_{i,j}n_{i+1,j}]
+ j \sum_j [-n_{k,j} - n_{k+1,j} + 2n_{k,j} n_{k+1,j} ] \ ,
\end{eqnarray}
\noindent where $E_0$ is the unperturbed energy of the $k=0$ phase,
$\Lambda \rightarrow \infty$, and for $K \rightarrow \infty$, we
may set $n_{0,j}=0$.  We write this Hamiltonian as
\begin{equation}
{\cal H}_B = E_0 + {\cal H}_0 + V_1 + V_2 + V_3 + V_4 \ ,
\end{equation}
where
\begin{equation}
\label{H0EQ}
{\cal H}_0 = \Delta \sum_i \sum_j n_{i,j} \ ,
\end{equation}
with $\Delta = 2J_0+j$,
\begin{equation}
\label{V1EQ}
V_1 = - \sum_i\sum_j j_{i,i+1} n_{i,j} n_{i+1,j}
- J_0 \sum_i \sum_{\langle j,j'\rangle } n_{i,j} n_{i,j'}
\end{equation}
\begin{equation}
V_2 = - (h/2) \sum_i \sum_j [a_{i,j}^\dagger + a_{i,j} ] \ ,
\end{equation}
\begin{equation}
V_3 = -j \sum_j (n_{k,j} + n_{k+1,j} ) \ ,
\end{equation}
\noindent where
\begin{equation}
j_{i,i+1} = \left\{ \begin{array} {ll}
j & \mbox{for } i \not= k\\
-j & \mbox{for } i=k
\end{array}
\right.
\end{equation}

\noindent and (with $\Lambda \rightarrow \infty$)
\begin{equation}
\label{V4EQ}
V_4 = \Lambda \sum_i \sum_j a_{i,j}^\dagger a_{i,j}^\dagger a_{i,j} a_{i,j} \ .
\end{equation}

\begin{figure}
\centerline{\psfig{figure=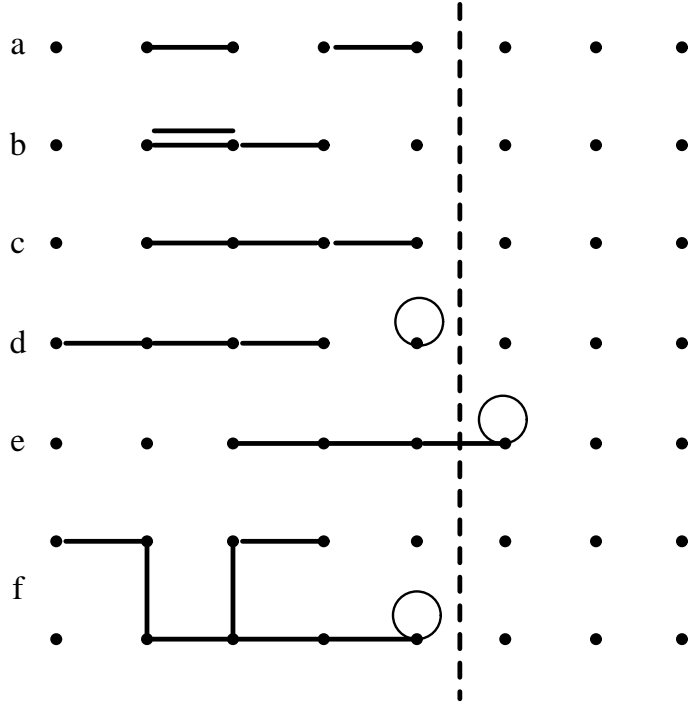}}
\vspace{0.5cm}
\caption{Representation of perturbative contributions to
the ground state
energy.  The dashed line corresponds to the location of the interface.
(a) A contribution which cannot occur.
(b) A term in which a given
$j$ is taken to second order. (c) The contribution to lowest
order in $j$ which involves four sites.
(d),(e),(f) Terms which involved factors of $h^2$.
(f) A term which involves the $J_0$ interaction.}
\label{f3}
\end{figure}

We now consider how the perturbative contributions to the energy depend
on the various coupling constants.  To carry out this discussion
it is convenient to introduce a diagrammatic representation of the
contributions to the perturbation expansion.  Each term of $V_1$
proportional to $j_{rs} n_r n_s$, where $r$ (and similarly $s$)
denotes a position label of the form $i,j$, is represented by a line
joining the two interacting sites $r$ and $s$ and $j_{rs}=J_0$
or $j_{rs}=j$ depending on whether sites $r$ and $s$ are in the same plane
or are in adjacent planes.  The perturbation in $V_2$
proportional to $a_r$ ($a_r^\dagger$) is represented by a minus
(plus) sign at the site $r$. However, for simplicity, since each site
involved with any of the
preceding interactions must be excited (i. e. must have both a "+" and
a "-" associated with it), we have not explicitly shown "+"'s and "-"'s
in Fig. \ref{f3}. The term in the perturbation $V_3$
proportional to $n_{k,j}$ ($n_{k+1,j}$) is represented by a
circle attached to the site $k,j$ ($k+1,j$).
Any term in perturbation theory which does not involve $V_4$
can be constructed from these elements.
Some simple examples
are shown in Fig. \ref{f3}.

We now define what we mean by connected terms.  Any term which involves
only a single site is connected.  Terms which involve more than one site
are connected only if all such sites are connected with respect to lines
representing terms of $V_1$.
If this is not the case, the term will be called disconnected.
Thus diagrams (a) and (d) of Fig. \ref{f3} are disconnected.

We now establish that
contributions from disconnected configurations of lines vanish.

Consider a disconnected diagram $\Gamma$ which consists of two
disjoint components, $\Gamma_A$ and $\Gamma_B$.  The contribution of
of this diagram is unchanged if we were to treat perturbatively the
system $S(\Gamma_A + \Gamma_B)$ in which all coupling constants $j_{rs}$
not in $\Gamma_A$ or $\Gamma_B$ are set to zero.  But because $\Gamma_A$
and $\Gamma_B$ are disjoint systems, we have $E(\Gamma_A + \Gamma_B)=
E(\Gamma_A) + E(\Gamma_B)$.  This result indicates that there are NO
disconnected terms
in the ground state energy which involve simultaneously an exchange
constant from one component $\Gamma_A$ and an exchange constant from
the other component $\Gamma_B$.
Thus disconnected diagrams can be omitted from further consideration.

\begin{figure}
\centerline{\psfig{figure=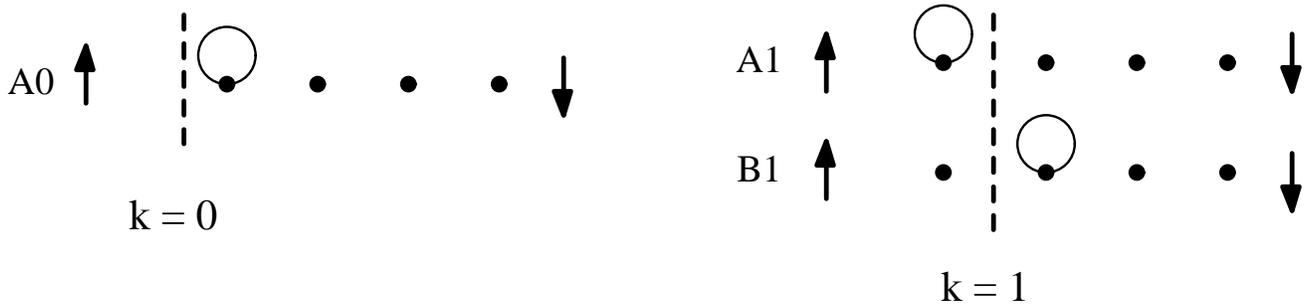}}

\vspace{0.5cm}
\caption{Perturbative contributions to the ground-state
energy used to
evaluate $\Delta E_0 = E_0 - E_1$.  Panel A0 corresponds to the case
when the interface is at $k=0$;  panels A1 and B1 to the case $k=1$.}
\label{f4}
\end{figure}

\begin{figure}
\centerline{\psfig{figure=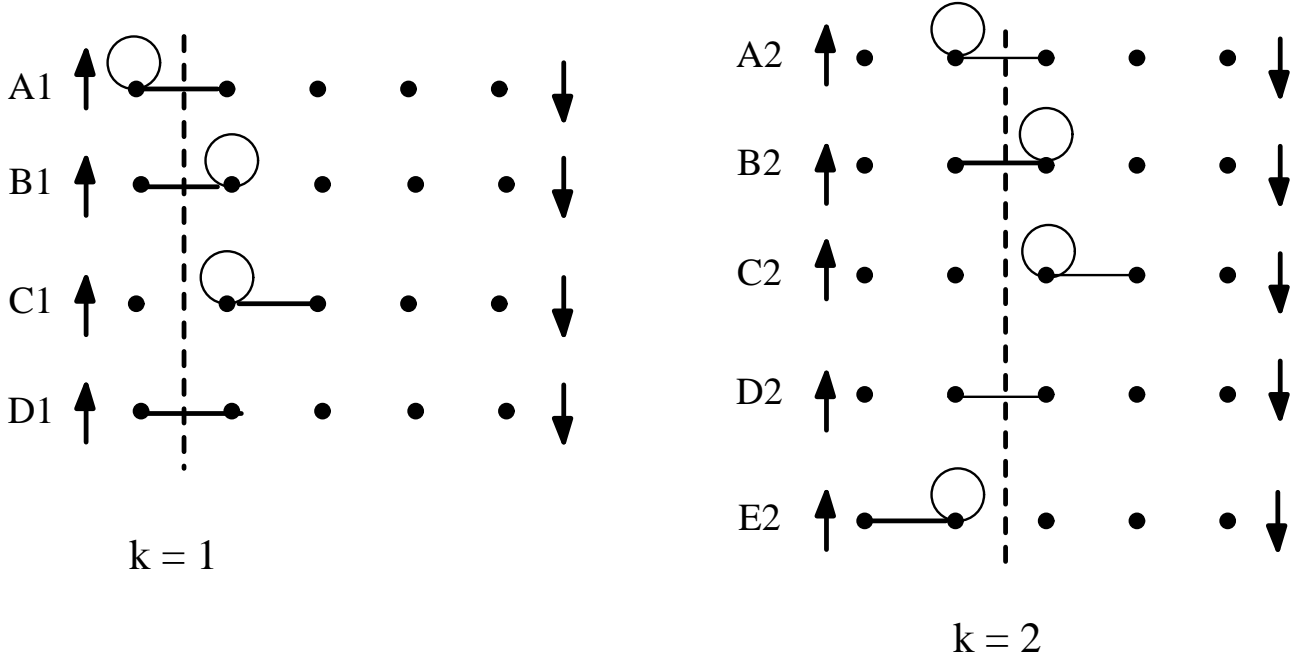}}

\vspace{0.5cm}
\caption{Perturbative contributions to the
ground-state energy used to
evaluate $\Delta E_1 = E_1 - E_2$.  Panels A1 -- D1 are for the case
when the interface is at $k=1$.  Panels A2 -- E2 correspond to
$k=2$.  The dashed line represents the position of the
interface.  Contributions from diagrams A1, B1, C1, and D1 are equal to
those from diagrams A2, B2, C2, and D2, respectively.  So
$\Delta E_1$ is determined by E2.}
\label{f5}
\end{figure}

To evaluate $\Delta E_k$ it is apparent from the form of Eq. (\ref{WALL})
that we only need to keep contributions which appear when the interface
is at position $k+1$ but NOT when it is at position $k$.
Note that if the diagram does not involve the interface
potential $V_3$, it has no dependence on position and will make
equal contributions in the two cases.
So $\Delta E_k$ has contributions from diagrams
which both a) involve $V_3$, the interface potential, and b) can occur
when the interface is at position $k+1$ but not at position $k$.
Such a connected diagram must involve sites in rows
1, 2, ... $k+1$.  This can be done with diagrams involving
$k$ or more lines.  As we will see in a moment, the
dominant contribution involves the least number of lines.
To take the least number of lines, means that we take
$k$ lines representing $j$ and no lines representing $J_0$.
Thus the dominant diagrams are linear chain diagrams, and we
may therefore consider a Hamiltonian in which the
index $j$ in Eqs. (\ref{H0EQ})--(\ref{V4EQ}) is omitted.
We illustrate the diagrams with the minimum number of lines
which contribute to $\Delta E_k$ for $k=0$ and $k=1$ in
Figs. \ref{f4} and \ref{f5} respectively.

To check that it is indeed the linear chain diagrams which give the
lowest order contribution to $\Delta E_k$
it is necessary to give a more detailed analysis of the
contribution of a diagram involving, say, $p$ different lines
representing $V_1$ and, as we explained, necessarily involving
at least one interface potential term $V_3$.
Note that both these perturbations, $V_1$ and $V_3$,
involve occupation numbers $n_r$, which vanish when there is no
excitation at site $r$.  For every site $r$ involved in a $V_1$ or
$V_3$ interaction, it is necessary to create an excitation
so that $n_r$ can be evaluated in a virtual state in which
$n_r=1$.  Subsequently, in order to get back into the ground
state we must destroy the excitation on the site $r$.
Thus in all a diagram involving $p$ different lines
will involve $p+1$ sites and therefore give rise to a perturbative
contribution to the energy which is of order $\delta_p E$, where
\begin{equation}
\delta_p E = h^{(2p+2)} j^{p_1} J_0^{p_2}(j/\Delta)/\Delta^{2p+1+p_1+p_2} \ ,
\end{equation}
where of the $p$ lines, $p_1$ are associated with $j$ and $p_2$ with
$J_0$ [see Eq. (\ref{V1EQ})].  In writing this equation we included
the factor $(j/\Delta)$ to take account of the necessary factor
of $V_3$.  At this point, it is clear that
to have a diagram which occurs for position $k+1$ but not for $k$
it is best to invoke a linear diagram, and not one which
reaches more than one row perpendicular to the interface.
Unnecessary factors of $J_0$ in $V_1$ will give rise to
additional factor of $J_0h^2/\Delta^3 \ll  1$.   So,
we conclude that to leading order
\begin{equation}
\Delta E_k = C_k J_0^2 [jh_0^2/J_0^3]^{k+1} \ ,
\end{equation}
where $C_k$ is a constant which must be determined by an
explicit calculation and to leading order we set $\Delta = 2J_0$.

Now let us carry out a detailed calculation for the
simplest case, namely for $k=0$.  From Fig. \ref{f4} we obtain
\begin{equation}
\label{RS}
\Delta E_0 = - E(A1) \ ,
\end{equation}
where $E(A1)$ is the contribution to the energy in diagram $A1$
of Fig. \ref{f4}.  Thus[\onlinecite{M66}]
\begin{eqnarray}
\Delta E_0 & = & - \langle 0| V_2 {1 \over {\cal E} } V_3
{1 \over {\cal E} } V_2 | 0 \rangle \nonumber \\
&=& - \langle 0 | (-h/2) a_1 {1 \over {\cal E} } (-j a_1^\dagger a_1)
 {1 \over {\cal E} } ( - h/2) a_1^\dagger | 0 \rangle \nonumber \\
&=& (h^2/4)(j/\Delta^2) = jh^2/(16J_0^2) \ .
\end{eqnarray}
Here and below the excitation energies $\cal E$ will be
$-r \Delta = -2rJ_0$, where $r$ is the number of excitations
in the virtual state.

Next we calculate $\Delta E_1 = -E(E2)$ from diagram $E2$ of
Fig. \ref{f5}.
Here we have to sum over the different orderings of the perturbations.
\begin{eqnarray}
\lefteqn{\Delta E_1 =} \nonumber \\
&& - \langle 0 | \biggl[
(-h/2)a_1 {1 \over {\cal E} } (-h/2) a_2
+ (-h/2)a_2 {1 \over {\cal E} } (-h/2) a_1 \biggr] {1 \over {\cal E}}
\biggl[ (-jn_1n_2 ){1 \over {\cal E}} (-jn_2) +
(-jn_2 ){1 \over {\cal E}} (-jn_1 n_2) \biggr] \nonumber \\
&& \ \ \times {1 \over {\cal E}}
\biggl[  (-h/2)a_1^\dagger {1 \over {\cal E} } (-h/2) a_2^\dagger
+ (-h/2)a_2^\dagger {1 \over {\cal E} } (-h/2) a_1^\dagger \biggr] |0 \rangle
\nonumber \\ && \ \ - \langle 0 | \biggl[
(-h/2)a_1 {1 \over {\cal E} } (-h/2) a_2
+ (-h/2)a_2 {1 \over {\cal E} } (-h/2) a_1 \biggr] {1 \over {\cal E}}
(-jn_1n_2 ){1 \over {\cal E}} (-h/2)a_1^\dagger {1 \over {\cal E} }
(-jn_2) {1 \over {\cal E}} (-h/2) a_2^\dagger |0 \rangle
\nonumber \\ && \ \ - \langle 0 |
(-h/2)a_2 {1 \over {\cal E} } (-jn_2) {1 \over {\cal E}}
(-h/2)a_1 {1 \over {\cal E}} (-jn_1n_2 ){1 \over {\cal E}}
\biggl[  (-h/2)a_1^\dagger {1 \over {\cal E} } (-h/2) a_2^\dagger
+ (-h/2)a_2^\dagger {1 \over {\cal E} } (-h/2) a_1^\dagger \biggr] |0
\rangle . \nonumber \\
\end{eqnarray}
When simplified this yields
\begin{equation}
\Delta E_1 = 2 (h/2)^4 j^2 / \Delta^5 = (1/256)j^2h^4/J_0^5 \ .
\end{equation}

\begin{figure}
\centerline{\psfig{figure=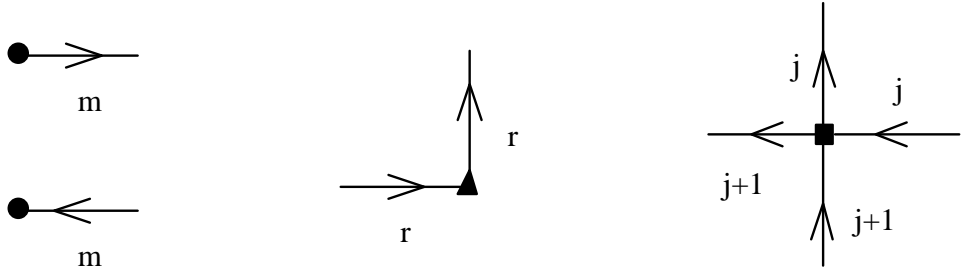}}
\vspace{0.5cm}
\caption{Vertices for Matsubara diagrams:
(a) transverse field vertex $-(h/2)(a_m^\dagger)$ (top) and
$-(h/2)a_m$ (bottom); (b) $-j\delta_{r,k}a_r^\dagger a_r$;
(c) $-j n_j n_{j+1}=ja_j^\dagger a_{j+1}^\dagger a_{j+1} a_j$.}
\label{f6}
\end{figure}

This calculation is hard to extend to $\Delta E_k$ for larger $k$
using naive perturbation
theory.  We need a more powerful formalism, namely Matsubara
diagrams[\onlinecite{MATSU}]. In this formalism one has
diagrams constructed from the following elements.  The perturbations
$V_1$, $V_2$, and $V_3$ are represented by vertices as shown in
Fig. \ref{f6}.  Each such vertex carries the appropriate factor
($(-h/2)$, $-j \delta_{r,k}$, and $-j$, respectively, where
$\delta_{r,k}$ is the Kronecker delta.  Also note that at leading
order $j_{i,i+1}=j$, because we never invoke the term $j_{k,k+1}$.).
Lines labeled with the same index are joined and a sum is taken
over all topologically inequivalent connected diagrams.  Each line
represents a Green's function $(z_\nu-\Delta)^{-1}$.  All
the indices are summed over.  The $z$'s are summed over the Matsubara
frequencies $z_\nu = 2 \nu \pi i /(k_BT)$ where $\nu$ runs over
all integers positive and negative.  One enforces
conservation of $z$, that is for each vertex the sum of all incoming
$z$'s minus the sum of all outgoing $z$'s must equal zero. For the
present case, this conservation law mean that at any vertex
which has only one line entering or leaving
the corresponding $z$ must be zero.  One can quickly see
that the $z$'s for all lines have to be zero.  So, in fact, there is
no sum over $z$ to be done.  (Such sums normally lead to
the Bose occupation number factors.)

\begin{figure}
\centerline{\psfig{figure=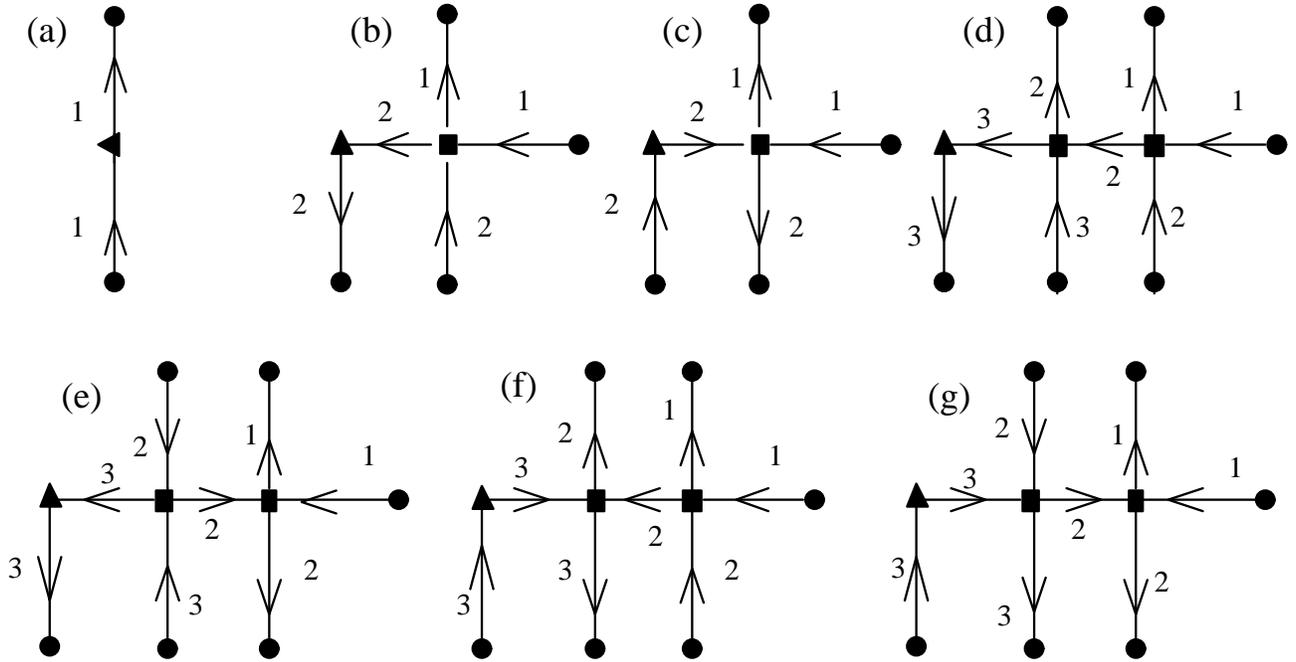,width=17cm}}
\vspace{0.5cm}
\caption{Matsubara diagrams.  The vertices are explained in
the preceding
figure.  The lines are labeled by an index $i$ to
represent the Green's function $(z-\Delta)^{-1}$.
This calculation is for a linear system with sites labeled 1, 2,
....  These diagram give $-\Delta E_m$ following the reasoning
of Figs. 4 and 5.  Diagram (a) is for $m=0$, diagrams (b) and (c)
are for $m=1$ and diagrams (d)--(g) are for $m=2$.  No new topology is
obtained by reversing the direction of the line labeled "1".
However, one  may independently reverse the direction of
all other lines
giving rise to a degeneracy $2^m$.}
\label{f7}
\end{figure}

{}From Fig. \ref{f7} we see that
\begin{equation}
\Delta E_0 = - E_a \ ,
\end{equation}
where $E_a$ is the energy from diagram (a) of Fig. \ref{f7}.  (A similar
notation for the other diagrams is used below.)  Thus
\begin{equation}
\Delta E_0 = - (-h/2) (-j) (-h/2) (-\Delta )^{-2} =
h^2j / (4 \Delta^2) = h^2j/(16J_0^2) \ ,
\end{equation}
as before.  To obtain this result, note that diagram (a) has two filled
circle vertices (each carrying a factor $-h/2$), one triangle
(carrying a factor $-j$), and two lines, each of which carries a
factor $(z-E_i)^{-1}=\Delta^{-1}$.
Also from Fig. \ref{f7} we have
\begin{equation}
\Delta E_1 = - (E_b + E_c) = -2E_b \ .
\end{equation}
Since diagram (b) has four $h$ vertices, two $j$ vertices, and five
lines, we have that
\begin{equation}
\Delta E_1 = -2 (-h/2)^4 (-j) (-j) (-\Delta)^{-5}
= h^4 j^2 / (8\Delta^5) = h^4j^2/(256J_0^5) \ .
\end{equation}
Finally, from Fig. \ref{f7} we have
\begin{equation}
\Delta E_2 = - (E_d + E_e + E_f + E_g) = -4E_d \ .
\end{equation}
Since diagram (d) has six $h$ vertices, three $j$ vertices, and eight lines
\begin{equation}
\Delta E_2 = -4 (-h/2)^6 (-j) (-j)^2 (-\Delta)^{-8}
= h^6 j^3 / (16\Delta^8) = h^6j^3/(4096J_0^8) \ .
\end{equation}
Evidently, the general result is
\begin{equation}
\Delta E_k = J_0 (jh^2/16J_0^3)^{k+1} \ .
\label{eqn:ep}
\end{equation}

\section{CONCLUSIONS}

\begin{figure}
\centerline{\psfig{figure=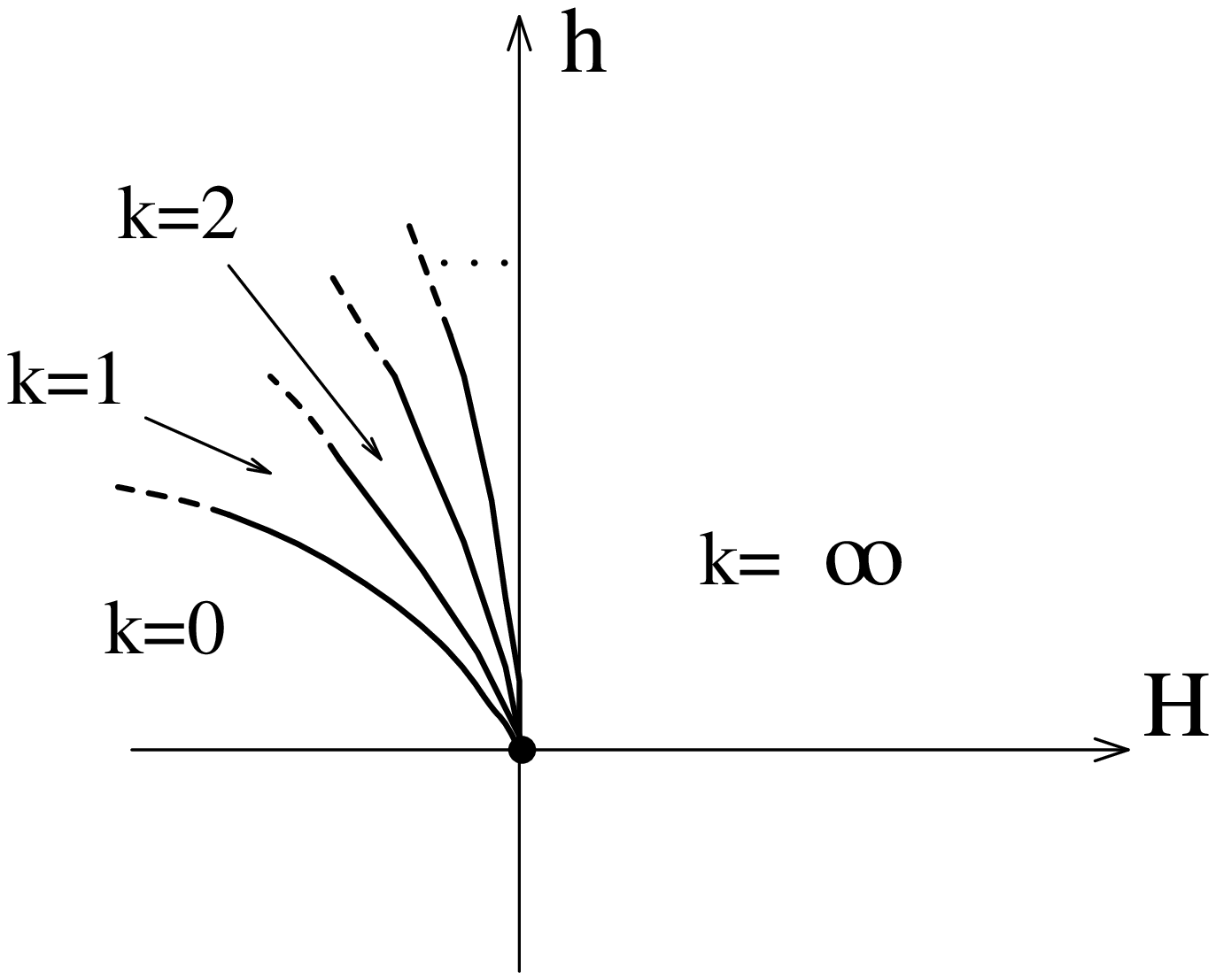,width=3.0in}}
\vspace{0.5cm}
\caption{Schematic representation of the phase diagram for interface
unbinding transitions in the transverse Ising model.}
\label{f8}
\end{figure}

Equation (\ref{eqn:ep}) indicates that the boundary between phases
with $k=p$ and $k=p+1$ is given to leading order by
$H_p^* = - J_0 (jh^2/16J_0^3)^{p+1}$. The resulting phase diagram is shown
schematically in Fig. \ref{f8}.

The analysis presented above was based on retaining only the
leading-order (in $h/J_0$ and $j/J_0$)
term in the surface--interface interaction. Although we cannot rule
out the possibility that the neglected higher-order interactions could
become dominant for very large $k$ (for fixed $h$ and $J_0$), we do
not expect to observe any qualitative corrections to the phase diagram
in this limit. This is because there are no competing interactions
which would make correlation functions oscillatory at large distance
and therefore it seems implausible that the positive sign of
$\Delta E_p$ can be changed by
the neglected higher-order terms[\onlinecite{FS87}].

To summarize we have shown that quantum fluctuations can stabilize an
infinite sequence of layering transitions in a three-dimensional
transverse Ising model.

\begin{acknowledgements}
We thank Malte Henkel for helpful discussions.
Work at the University of Pennsylvania was partially supported by
the National Science Foundation under Grant No. 95-20175. JMY
acknowledges support from the EPSRC and CM from the EPSRC and the
Fondazione ``A. della Riccia'', Firenze.
\end{acknowledgements}

\newpage

\vspace{0.5cm}

\newpage

\vspace{0.5cm}

\newpage

\vspace{0.5cm}
\newpage

\vspace{0.5cm}

\begin{references}

\bibitem{unbinding} S. Dietrich in {\em Phase Transitions and Critical
Phenomena}, edited by C. Domb and J. L. Lebowitz (Academic, New York,
1988), Vol. 12. p. 1 ; R. Pandit, M. Schick and M. Worthis, {\em
Phys. Rev. B} {\bf 26}, 5112 (1982)


\bibitem{DG78}
M. J. De Oliveira and R. B. Griffiths, Surf. Sci. {\bf 71}, 687
(1978).

\bibitem{DY85}
P. M. Duxbury and J. M. Yeomans, J. Phys. A {\bf 18}, L983 (1985).

\bibitem{RS73} R. B. Stinchcombe, J. Phys. C {\bf 6}, 2459 (1973).

\bibitem{dG63} P.G. de Gennes, Solid St. Commun. {\bf 1}, 132 (1963).
\bibitem{HH95}
M. Henkel, A.B. Harris and M. Cieplak, Phys. Rev. B{\bf 52}, 4371
(1995).

\bibitem{Selke} W. Selke, in {\em Phase Transitions and Critical
Phenomena}, edited by C. Domb and J. L. Lebowitz (Academic, New York,
1992), Vol. 15.

\bibitem{HM95}
A.B. Harris, C. Micheletti and J.M. Yeomans, Phys. Rev. Lett. {\bf
74}, 3045 (1995); Phys. Rev. {\bf B52}, 6684 (1995).

\bibitem{CLH}
C. L. Henley, Phys. Rev. Lett. {\bf 62}, 2056 (1989).

\bibitem{EFS}
E. F. Shender, Sov. Phys. JETP {\bf 56}, 178 (1982).

\bibitem{JV}
J. Villain and M. B. Gordon, J. Phys. C {\bf 13}, 3117 (1980).

\bibitem{M66}
A. Messiah {\em ``Quantum Mechanics''}, North-Holland, Amsterdam
(1966).

\bibitem{MATSU}J. W. Negele and H. Orland {\em `` Quantum
Many-Particle Systems''}, Addison-Wesley (1988)


\bibitem{FS87}
M. E. Fisher and A. M. Szpilka, Phys Rev. B {\bf 36}, 644 (1987).












\end{references}
\end{document}